\documentclass[aps,prl,twocolumn,nobibnotes]{revtex4-1}
\usepackage{amsmath,amssymb,amsfonts,amsbsy}
\usepackage{color}
\usepackage{bm}
\usepackage{graphicx}
\usepackage[a4paper,vmargin={20mm,20mm},hmargin={20mm,10mm}]{geometry}
\usepackage{subfigure}
\usepackage{paralist}
\usepackage{bbold}
\usepackage{braket}
\usepackage{ctable}
\usepackage{caption}
\usepackage{wrapfig}
\usepackage{array}
\usepackage{dcolumn}
\usepackage{multirow}
\usepackage{subfig}
\usepackage{float}
\usepackage[version=3]{mhchem}
\usepackage{bbm}
\usepackage{algorithm}
\usepackage{algorithmicx}
\usepackage{algpseudocode}

\begin{document}
\title{
    A Projector Quantum Monte Carlo Method for non-linear wavefunctions
}
\author{Lauretta~R.~Schwarz}
\email{lrs37@cam.ac.uk}
\affiliation{University of Cambridge, Lensfield Road, Cambridge CB2 1EW, U.K.}
\author{A.~Alavi}
\email{A.Alavi@fkf.mpg.de}
\affiliation{Max Planck Institute for Solid State Research, Heisenbergstra{\ss}e 1, 70569 Stuttgart, Germany}
\affiliation{University of Cambridge, Lensfield Road, Cambridge CB2 1EW, U.K.}
\author{George~H.~Booth}
\email{george.booth@kcl.ac.uk}
\affiliation{Department of Physics, King's College London, Strand, London, WC2R 2LS, U.K.}

\date{\today}

\begin{abstract}
We reformulate the projected imaginary-time evolution of Full Configuration Interaction Quantum Monte Carlo in terms of a Lagrangian minimization.
This naturally leads to the admission of polynomial complex wavefunction parameterizations, circumventing the exponential scaling of the approach.
While previously these functions have traditionally inhabited the domain of Variational 
Monte Carlo, we consider recent developments for the identification of deep-learning neural networks to
optimize this Lagrangian, which can be written as a modification of the propagator for the wavefunction dynamics.
We demonstrate this approach with a form of Tensor Network State, and use it to find solutions to the strongly-correlated Hubbard model, as well as
its application to a fully periodic {\em ab-initio} Graphene sheet. The number of variables which can be simultaneously optimized 
greatly exceeds alternative formulations of Variational Monte Carlo, allowing for systematic improvability of the 
wavefunction flexibility towards exactness for a number of different forms,
whilst blurring the line between traditional Variational and Projector quantum Monte Carlo approaches.
\end{abstract}

\maketitle

The description of quantum many-body states in strongly-correlated 
systems is central to understanding a wealth of complex emergent 
phenomena in condensed matter physics and quantum chemistry. 
The problem is well defined; the Hamiltonian is known, and the solution
is a linear superposition of all possible classical configurations of particles.
However, this conceals exponential complexity in the wavefunction which in general prohibits
both the storage and manipulation of these linear coefficients. 

To deal with this exponentially large Hilbert space, one approach is to sample the space stochastically. 
For studies of the ground state of quantum systems, this is broadly split into two separate categories,
{\it Projector} and {\it Variational} Monte Carlo (PMC / VMC)\cite{booknightingaleumrigar,RevModPhys.73.33}. 
In the first, an operator written as a decaying function of the Hamiltonian is
continually applied to a stochastic representation of the full wavefunction. This projects 
out the higher energy components, leaving a stochastic sampling of the dominant, (generally ground-state) eigenfunction. 
By contrast, in VMC a compact, polynomial-complex approximate wavefunction ansatz is imposed, generally with a small number of variational 
parameters. State-of-the-art methods to optimize this wavefunction then involve sampling and accumulating the gradient and hessian of the energy 
with respect to the parameters in the tangent space of the current wavefunction. This is done by projecting into and sampling from the 
exponential configurational space. Once a stochastic representation of these quantities is obtained, updates to 
the wavefunction parameters are found by a variety of iterative techniques until convergence of this non-linear parameterization is achieved.

One promising emerging technique is Full Configuration Interaction Quantum Monte Carlo (FCIQMC), a projector quantum Monte Carlo
which stochastically samples both the wavefunction and the propagator in Fock space\cite{2009JChPh.13193710,solids_Nature13}. 
By exploiting sparsity
inherent in the wavefunction of many representations of quantum systems, essentially exact results can 
be obtained with only small fractions of the Hilbert space occupied at any one time. However, despite often admitting 
highly accurate solutions for systems far out of reach of many alternative approaches, the method is formally exponentially scaling with system size, albeit often weakly. 
In order to advance to larger and condensed phase systems, one approach is to exploit the fact that electron correlation is, in general, inherently local.
Two-point correlation functions (away from criticality) will decay exponentially with distance, whilst the screening of the Coulomb interaction
in bulk systems will result in local entanglement of nearby electrons, with distant electrons behaving increasingly independently\cite{RevModPhys.82.277}. 

Following in the success of the FCIQMC approach for finite quantum systems, we aim to extend it to exploit this
locality, to formally contain the scaling to polynomial cost. 
This is done by imposing a non-linear, yet systematically improvable ansatz of the form of a Correlator Product 
State (CPS), which explicitly correlates plaquettes of locally neighbouring degrees of freedom\cite{PhysRevB.80.245116,PhysRevB.86.064402}. 
Related wavefunctions have also 
been called Entangled Plaquette States or Complete Graph Tensor Networks to stress their connection to higher-dimension
generalizations of matrix product states\cite{1367-2630-11-8-083026,1367-2630-12-10-103039,1367-2630-12-10-103008,C0CP01883J}. 
In formulating this, we develop connections between Projector and Variational
quantum Monte Carlo, and propose new methodology for the optimization of arbitrary non-linear wavefunction parameterization. 
This approach is shown to confer a number of benefits compared to
state-of-the-art wavefunction optimization\cite{2007JChPh.126h4102T,PhysRevB.71.241103,2007JChPh.127a4105S,2007JChPh.127a4105S,PhysRevB.85.045103,2017arXiv1702}.
The number of parameters which can be handled even brings into scope more sophisticated wavefunctions, including other tensor network parameterizations\cite{PhysRevLett.99.220602,PhysRevB.91.165113}, while the
Lagrangian can also be formulated for non-linear constraints in alternative applications. 
We apply this approach to a number of model and {\em ab-initio} systems, showing that systematic improvability and exceedingly large parameter spaces
can be handled for these complex optimization problems.

The CPS wavefunction defines `correlators' as diagonal operators (to optimize) which directly encode the entanglement within 
sets of single-particle states (which in this work are exclusively neighbouring), as
$\hat{C}_{\lambda} = \sum_{{\bf n_{\lambda}}} C_{{\bf n_{\lambda}}} 
 \hat{P}_{{\bf n_{\lambda}}} $,
where $\hat{P}_{{\bf n_{\lambda}}} = \ket{{\bf n_{\lambda}}}\bra{{\bf n_{\lambda}}} $
is the projection operator for the set of {\em all} many-body Fock states ${\bf n_{\lambda}}$ in the 
correlator $\lambda$, with adjustable amplitudes $C_{{\bf n_{\lambda}}}$.
The CPS is then written as a multi-linear product of 
correlators acting on a chosen reference state, $\ket{\Phi}$. In this work, this reference state is a single Slater determinant (which can also be variationally optimized), but other 
reference states are possible\cite{AGP_Neuscamman,AGP_Sorella}. The final CPS wavefunction is therefore represented as
$\ket{\Psi_{\mathrm{CPS}}} = \prod_{\lambda} \hat{C}_{\lambda} \ket{\Phi}$.
It can be shown that a number of different phases and wavefunctions 
can be expressed in this form, including RVB and Laughlin wave functions\cite{PhysRevB.80.245116}. 
As the number of degrees of freedom in the system grows, the complexity of the 
wavefunction grows only linearly.
Additionally, this choice of low-rank factorization of the wavefunction is systematically improvable 
in the limit of increasing 
correlator size as it recovers longer-ranged entanglement effects, but this admits many variables to optimize. 
VMC techniques have been used previously for similar tensor network forms, but the growth of parameters has led to limited success in 
recovering long-range entanglement or thermodynamic limit 
results\cite{PhysRevLett.99.220602,PhysRevB.91.165113}. We now consider a new, robust and efficient approach to
handle these many parameters, derived in part from the FCIQMC approach, which can be considered as the limit of a single large correlator.

\textit{Combining PMC and VMC.--} The FCIQMC (and some other PMC\cite{PhysRevLett.95.100201}) methods are simulated through stochastic dynamics given by
\begin{eqnarray}
|\Psi_0\rangle = \lim_{k \rightarrow \infty} (1 - \tau ({\hat H} - {\hat I} E_0))^k | \psi^{(0)} \rangle	, \label{eqn:FCIQMCDyn}
\end{eqnarray}
with $\tau$ chosen to be sufficiently small, where $\Psi_0$ is the ground state of the system, and $E_0$ is the self-consistently obtained ground state energy\cite{2009JChPh.13193710}.
This can be considered both as a first-order approximation to imaginary time dynamics as $e^{- \beta {\hat H}}|\psi^{(0)} \rangle$, or as a power
method to project out the dominant, lowest energy eigenvector of ${\hat H}$\cite{2015JChPh14933112}. Alternatively, a VMC perspective considers finding the
variational minimum of the Ritz functional, $\frac{\langle \Psi | {\hat H} | \Psi \rangle}{\langle \Psi | \Psi \rangle}$, through optimization of the wavefunction parameters.

These approaches can be shown to be analogous by considering the minimization of a positive-definite Lagrangian,
\begin{eqnarray}
 \mathcal{L} \left[ \Psi (Z_{\sigma})\right] = \braket{\Psi|\hat{H}|\Psi} - E_0 \left( 
 \braket{\Psi|\hat{I}|\Psi} - A \right),	\label{eqn:Lagrangian}
\end{eqnarray}
where normalization (up to an arbitrary constant $A$) is enforced by a Lagrange multiplier,
which at convergence is given by $E_0$. It is simple to show that the minimum of this
functional is the same as that given by the Ritz functional. We can 
consider a simple gradient descent
minimization of all variational parameters, $\{ Z_{\sigma} \}$ in 
Eq.~\ref{eqn:Lagrangian}, with step size $\tau_k$, as
\begin{eqnarray} 
 Z_{\sigma}^{(k+1)} &=& Z_{\sigma}^{(k)} - \tau_{k} \frac{\partial \mathcal{L}  
 \left[ \Psi^{(k)} \right]}{\partial Z_{\sigma}}	. \label{eqn:GD} 
\end{eqnarray}
Projecting the equations into the full Hilbert space of configurations, $\{ |\bf{m}\rangle \}$, we obtain
\begin{eqnarray} 
 Z_{\sigma}^{(k+1)} &=& Z_{\sigma}^{(k)} - \tau_{k} \sum_{{\bf nm}} \langle \frac{\partial 
 \Psi^{(k)}}{\partial Z_{\sigma}}|{\bf m}\rangle (H_{{\bf mn}}-E^{(k)} 
 \delta_{{\bf mn}}) \langle {\bf n} | \Psi^{(k)} \rangle . \label{eqn:Update}
\end{eqnarray}
If the chosen wave function is an expansion of linearly independent configurations, then
this will return exactly the `imaginary-time' dynamics of Eq.~\ref{eqn:FCIQMCDyn} and the FCIQMC
master equations, demonstrating the deep connection between imaginary-time propagation, gradient descent and
the power method\cite{2014arXiv1405.4980B}.

However, here we aim to go beyond this. In keeping with 
FCIQMC, the summations are replaced by random samples of both the wavefunction and Hamiltonian connections.
The sum over ${\{\bf n}\}$ is stochastically sampled via a Metropolis Markov chain, to evaluate a stochastic representation of the
wavefunction\cite{1953JChPh.11699114,2015JChPh14933112,HASTINGS01041970,
PhysRevE.83.066706,2010JChPh.133q4120T}. Each iteration consists of 100,000-200,000 random samples of the wavefunction
for the largest results shown.
Similarly, a small selection of configurations, $\{ {\bf m} \}$, are 
sampled from the set of non-zero connections via $H_{{\bf mn}}$
in the manner of FCIQMC, and unbiasing for the probability with a computed 
normalized generation probability\cite{2014MolPh.112.1855B,heatbath}. 
Furthermore, the derivatives $\langle \frac{\partial \Psi^{(k)}}{\partial 
Z_{\sigma}}|{\bf m}\rangle$ can be efficiently evaluated from the 
respective wavefunction amplitudes  
$\langle \Psi^{(k)}|{\bf m}\rangle$. Technical details on 
the sampling of this gradient can be found in the supplementary material.

This stochastic gradient descent (SGD) of the Lagrangian results in an iteration cost that is independent of the 
size of the Hilbert space and thus renders this methods inherently suitable for 
large scale systems. 
It also admits a number of advantages over state-of-the-art VMC optimization\cite{2007JChPh.126h4102T,PhysRevB.71.241103,2007JChPh.127a4105S},
such as the avoidance of the construction of matrices in the tangent space of the wavefunction, whose sampling and manipulation becomes a
bottleneck for large numbers of parameters. 
Whilst Krylov subspace techniques have been proposed to circumvent this by projecting down to more manageable spaces\cite{PhysRevB.85.045103}, 
ill-conditioning can limit the efficiency of this approach\cite{2017arXiv1702}. 
Furthermore, diagonalization of the randomly sampled matrices required in some optimizations can lead to 
biases in the final parameters\cite{PhysRevLett.115.050603,doi:10.1021/acs.jctc.6b00480}.
Our approach also bears similarities with the Stochastic Reconfiguration method (SR)\cite{PhysRevB.71.241103,2007JChPh.127a4105S}, which can
also be considered an imaginary time propagation that differs from steepest descent in its definition of the metric in parameter space for the updates\cite{casula-2004}.
Due to this, SR requires projection of the equations into the fixed tangent space of the current wavefunction and stabilization of the resultant
matrix equations\cite{2007JChPh.127a4105S}.
However, the proposed matrix-free stochastic application of Eq.~\ref{eqn:GD} describes a
quasi-continuous optimization, where the error bar at convergence represents both the stochastic error in the sampling, and the variation in the 
wavefunction as it is sampled.
In addition, the dynamic also provides a straightforward route to unbiased computation of 
the two-body reduced density matrix\cite{Wagner2013,2014JChPh.141x4117O}, $\Gamma_{pq,rs} = \braket{\Psi |a_{p}^{\dagger} a_{q}^{\dagger} 
 a_{s}a_{r}|\Psi}$. 
By evaluating $\braket{Q} = \mathrm{Tr} \left[ \Gamma \hat{Q} \right]$, arbitrary $1$- and $2$-body static properties can be found.
This includes the energy, spin and magnetic properties which here are computed in the results from the density matrix, rather than from the local energy as is
commonly performed in VMC.

However, similar SGD approaches have been considered before with little success for
large numbers of variables, due to the slow (linear) convergence of the parameters 
as $\mathcal{O} \left( \frac{1}{k} + \frac{\sigma}{\sqrt{k}} \right)$ where $\sigma$ is the 
variance in the gradient\cite{PhysRevLett.79.1173,robbins1951}. Improving on this to obtain the convergence rate
of state-of-the-art quasi-second order methods involves advances in SGD methods, used in
the field of deep learning algorithms of neural networks\cite{NeuralNetworks,PhysRevLett.117.130501}. 
Analogously, these networks represent a flexible non-linear function with parameters to be optimized via minimization of a cost function,
often achieved via SGD schemes, similar to the one in Eq.~\ref{eqn:GD}\cite{939002,Jacobs1988295}.

The convergence can be accelerated via the addition of a `momentum', whereby the update depends on not just the
current iterate, but retains a memory of the one before. Propagation then results in the accumulation of velocity in the direction of persistent decrease
in energy, thereby accelerating the update in directions of low curvature over multiple iterations\cite{Qian1999145}, 
formally accelerating the convergence rate to a second-order $\mathcal{O} \left( \frac{1}{k^{2}} + \frac{\sigma}{\sqrt{k}} \right)$. Mathematically, the 
stochastic projection is given by a monic polynomial of the propagator of degree $k$, such that $ {\bf \Psi}^{(k)} = p^{k}_{A} \left( {\bf A}\right) {\bf \Psi}^{(0)}$. In
the SGD scheme of Eq.~\ref{eqn:FCIQMCDyn}, this is a simple polynomial of ${\bf A}^k$, akin to the power method. However, the optimal projection will
be a polynomial approximation to a function whose value at the desired eigenvalue of the propagator is one, and whose maximum absolute
value in the range of the rest of the spectrum is minimized. This is best represented by using a shifted and scaled Chebyshev polynomial approximation
to the projection. The success of the Lanczos approach as a second-order optimization, as well as other 
deterministic projections can also be rationalized in this fashion\cite{LanczosBook,Evangelista2016}.
 
An optimal version of this projector can be formulated as Nesterov's accelerated approach\cite{nesterov1983method}, whereby the 
sequence $\lambda_{0} = 0$, $\lambda_{k} = \frac{1}{2}+ \frac{1}{2}\sqrt{1+4\lambda^{2}_{k-1}}$, $\gamma_{k} = \frac{1-\lambda_{k}}{\lambda_{k+1}}$ 
is defined 
and starting at an initial point $Z_{\sigma}^{(1)} = Y_{\sigma}^{(1)}$, the algorithm stochastically iterates the equations\cite{beck2009fast},
 \begin{eqnarray} \label{nesterovseries}
 Y_{\sigma}^{(k+1)} &=& Z_{\sigma}^{(k)} - \tau_{k} \frac{\partial \mathcal{L} 
 \left[ \Psi^{(k)} \right]}{\partial Z_{\sigma}} \\
 Z_{\sigma}^{(k+1)} &=& \left( 1 - \gamma_{k} \right) Y_{\sigma}^{(k+1)} 
 + \gamma_{k} Y_{\sigma}^{(k)},
\end{eqnarray}
for $k \geq 1$.
While an optimal projection overall, this is no longer a gradient descent scheme, and as such there is no requirement
that each iteration will decrease the energy, and instabilities can be observed\cite{NIPS2014_5322,5173518}.
To mitigate this behaviour, we have found it beneficial to include a damping for the momentum, $d$, as
$\gamma_{k} \rightarrow \gamma_{k} e^{-\frac{1}{d}\left( k-1\right)}$.\cite{NIPS2014_5322,ODonoghue2013}
With a suitably chosen damping parameter the rate of convergence of the optimisation should not be hindered, since this is dominated in
the latter stages by the $\frac{\sigma}{\sqrt{k}}$ term for both 
accelerated and conventional gradient descent\cite{icml2013_sutskever13}.

The remaining arbitrariness concerns the step size (or `learning rate') $\tau_{k}$, which is crucial for the efficiency of the optimization.
Whilst decreasing the step size generally improves robustness, it slows convergence and increases autocorrelation time\cite{939002,Jacobs1988295}.
We found optimal convergence and accuracy achieved with a deep-learning technique denoted 
RMSprop\cite{RMSProp},  
an adaptive step size method which dynamically estimates an individual and  
independent $\tau^{(k)}_{Z_{\sigma}}$ for each parameter. This gives $\tau^{(k)}_{Z_{\sigma}} = \eta \left( \text{RMS} \left[ 
 g_{Z_{\sigma}} \right]^{(k)} \right)^{-1}$, where $\eta$ is a global parameter for all variables, and 
 $\text{RMS} \left[ g_{Z_{\sigma}} \right]^{(k)}$ represents the root mean 
square (RMS) of previous gradients for the variable up to 
the current iteration, $\text{RMS} 
\left[ g_{Z_{\sigma}} \right]^{(k)} = \sqrt{E \left[ g^{2}_{Z_{\sigma}} \right] 
 + \epsilon}$, evaluated by accumulating an 
exponentially decaying average of the squared gradients of the Lagrangian, $g$,
$E \left[ g^{2}_{Z_{\sigma}} \right]^{(k)} = \rho E \left[ 
 g^{2}_{Z_{\sigma}} \right]^{(k-1)} + \left( 1 -\rho \right) 
 g^{2}_{Z_{\sigma}}$. The small constant $\epsilon$ is added to better 
 condition the denominator 
 and $\rho$ is the decay constant. This dynamically 
adaptive, parameter-specific step-size, acts much like a preconditioner for the system, and allows 
the optimisation to take larger steps for those parameters with small and consistent
gradients, and vice versa.
This ensures robustness of the algorithm to large 
changes in gradients due to the stochastic nature of the gradient evaluation. 

\begin{figure}[htbp!]
\vspace{-10pt}
 \includegraphics[width=\columnwidth,trim=0.0cm 0.0cm 0.0cm 0.0cm,
 clip=true,keepaspectratio=true]{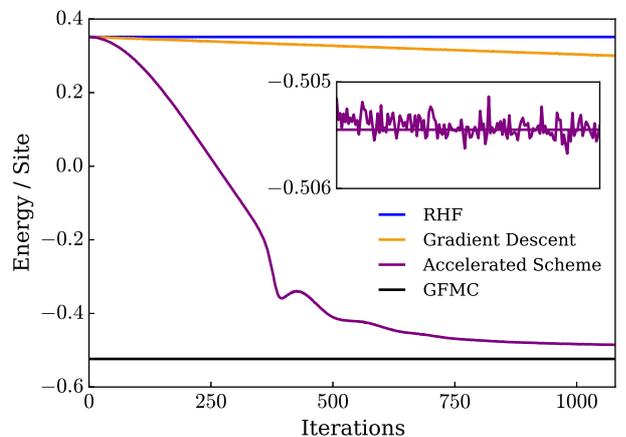}
 \vspace{-25pt}
 \caption{\footnotesize Convergence of CPS with $\mathcal{O}[10^5]$ parameters for SGD and 
 accelerated scheme with RMSProp algorithm for the $98$-site (tilted)
 $2$D Hubbard model at $U=8t$. GFMC energy is taken from Ref.~\onlinecite{PhysRevX.5.041041}.
 Inset shows fluctuations both in the statistical sampling of expectation values, and in the variation of the
 parameters.}
 \label{fig:convergence}
 \vspace{-15pt}
\end{figure}

\textit{Results.--} The demonstration of the ability of the algorithm to converge wavefunctions with many
parameters is shown in fig.~\ref{fig:convergence}, which considers a 98-site 2D Hubbard model at half-filling, with $U/t=8$.
In this study,
independent, overlapping five-site correlators centred on every site in the lattice were chosen to correlate with nearest neighbours,
allowing up to ten-electron short-ranged correlation to be directly captured, as well as long range correlation and symmetry-breaking through 
coupling between the overlapping correlators and the optimization of the Slater determinant.
The lattice and tiling of these correlator plaquettes is depicted in the supplementary materials.
Accurate results for this system 
are given by Greens-function Monte Carlo (GFMC)\cite{PhysRevX.5.041041}. Our CPS captures 97.9\% of the correlation energy of GFMC, 
with the remaining likely to be due to the lack of direct long-range two-body correlation. 
However, this parameterization still requires the simultaneous optimization of 
over $10^5$ parameters, beyond the capabilities of most VMC implementations, 
and demonstrates a striking advance in the rate of convergence afforded by the accelerated algorithm.

\begin{figure}[htbp!]
\vspace{-10pt}
 \includegraphics[width=\columnwidth,trim=0.0cm 0.0cm 0.0cm 0.0cm,
 clip=true,keepaspectratio=true]{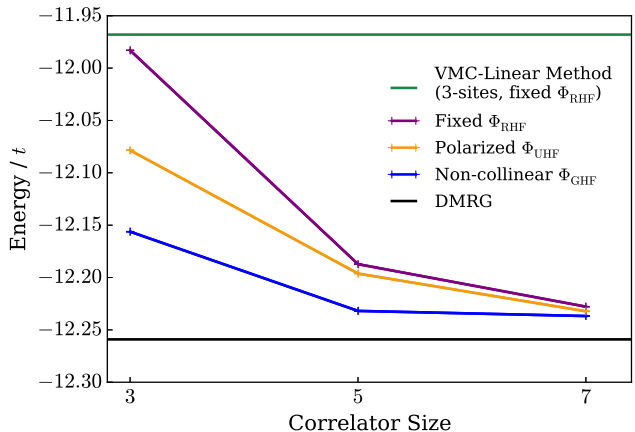}
 \vspace{-25pt}
 \caption{\footnotesize Convergence of energy for a range of 
 $\Psi_{\mathrm{CPS}}$ for $1\times 22$ Hubbard model. VMC Linear 
 Method and DMRG energies are taken from Ref.\cite{PhysRevB.84.205132}. Error bars are too small to be visible.}
 \label{fig:convergencesize}
 \vspace{-10pt}
\end{figure}

To consider the systematic improvability of the CPS ansatze, we consider the 1D, 22-site Hubbard model
with open boundary conditions, such that benchmark data can be found from the Density Matrix Renormalization Group (DMRG), which can be made numerically exact for this 1D system\cite{PhysRevB.84.205132}.
Results at half filling and $U = 4t$ are shown in fig.~\ref{fig:convergencesize}. For a wavefunction of three-site overlapping correlators 
and a fixed, non-interacting reference, we find a variationally lower result than previously published
for an identical parameterization via the state-of-the-art Linear Method optimization\cite{PhysRevB.84.205132,2007JChPh.126h4102T}.
This could be due to the bias from the non-linear operations (diagonalization) of random variables present in
these alternate algorithms\cite{PhysRevLett.115.050603,doi:10.1021/acs.jctc.6b00480}. 
We also investigate how increasing the size of the correlators in order to {\em directly} capture longer-ranged many-body correlation,
as well as optimizing spin-polarized ($\Phi_{UHF}$) or non-collinear ($\Phi_{GHF}$) Slater determinants rather than a paramagnetic orbital component ($\Phi_{RHF}$)
affects the quality of the wavefunction. The increased flexibility of this democratic wavefunction gives rise to systematic convergence towards DMRG 
with very small errorbars, despite requiring over quarter of a million variables.

\begin{figure}[htbp!]
\vspace{-10pt}
 \includegraphics[width=\columnwidth,trim=0.0cm 0.0cm 0.0cm 0.0cm,
 clip=true,keepaspectratio=true]{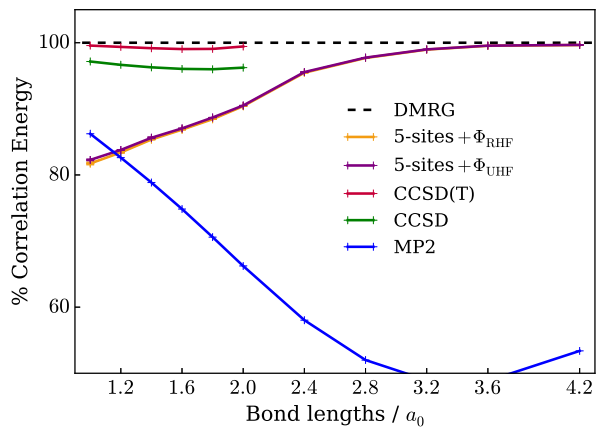}
 \vspace{-30pt}
 \caption{\footnotesize Percentage of DMRG correlation energy
 captured by $\Psi_{\mathrm{CPS}}$ for the symmetric dissociation of a linear chain of 50 hydrogen atoms. 
 Numerically exact DMRG, as well as high-level correlated quantum chemical methods of M{\o}ller-Plesset perturbation theory (MP2), coupled-cluster up to double excitations (CCSD) and with perturbative triple excitations (CCSD(T)) are included, with values taken from Ref.\cite{2006JChemPhys1.2345196}. The
 largest deviation in the total energy compared to DMRG across all bond lengths shown is 1.1kcal/mol per atom.}
 \label{fig:bindingcurve}
 \vspace{-10pt}
\end{figure}

{\it Ab-initio} systems can also be well treated in the same vein; stochastically sampling from both the 
configuration space of the wavefunction and from its $\mathcal{O}[N^4]$ connected configurations in Eq.~\ref{eqn:Update}, which are now far larger than 
found in the Hubbard model due to long-range interactions. 
We consider the symmetric dissociation of H$_{50}$ in a STO-6G basis\cite{hehre1969self}, a molecular model for strongly-correlated systems and a 
non-trivial benchmark system. This system has been treated not only with conventional quantum chemistry 
methods such as Coupled Cluster (CC) (which fail to converge at stretched bond-lengths beyond $2.0a_0$)\cite{2006JChemPhys1.2345196},
but also strongly-correlated approaches including DMFT and other embedding methods\cite{tsuchimochi2009strong,PhysRevB.89.201106,
PhysRevLett.106.096402}, due to the availability of numerically exact DMRG values for comparison\cite{2006JChemPhys1.2345196}. 
We parameterise our CPS with 
$5$-atom overlapping correlators, and both a fixed unpolarized reference, or stochastically optimised 
unrestricted reference determinant. 
At stretched bond lengths, nearly all of the DMRG correlation energy is captured, as the correlation length spans few
atoms, and on-site repulsion dominates. However, as 
the bond length decreases, a successively smaller percentage of the 
DMRG correlation energy is captured, as the entanglement of the electrons span larger numbers of atoms,
as can also be seen in the larger bond dimension required of DMRG at these geometries\cite{2006JChemPhys1.2345196}. 
Despite this, the correlation energy is so small at these lengths, that the maximum error in the total energy is only 1.1kcal/mol 
per atom, achieving chemical accuracy for the stretching of this system.

Fully periodic localized orbitals can also be used to construct a Fock space in which to form a CPS, and here we consider an infinitely 
periodic graphene sheet with $4 \times 4$ $k$-point sampling\cite{SolidsGaussianBasis}. From a double-zeta periodic Gaussian
basis, we choose one localized, translationally invariant $2p_z$ orbital centred on each carbon atom. Overlapping correlators consisting of the atoms on each hexagonal 
six-membered ring can then be constructed and the full Hamiltonian projected into this low-energy space, including a potential from the core electrons at the 
Hartree--Fock level\cite{StocCASSCF}. A generalized reference determinant is then stochastically optimized along with the correlators, giving a wavefunction 
parameterization of 67,584 parameters -- we believe the largest number of non-linear parameters for an {\em ab-initio} system to date. This is equivalent to a 
quantum chemical calculation of a complete active space of 32 orbitals, which is beyond that which could be treated by conventional techniques. This spans
the dominant strong correlation effects, but precludes high-energy many-body dynamic correlation and screening.

\begin{figure}[htbp!]
 \vspace{-5pt}
 \includegraphics[width=\columnwidth,trim=0.0cm 0.0cm 0.0cm 0.0cm,
 clip=true,keepaspectratio=true]{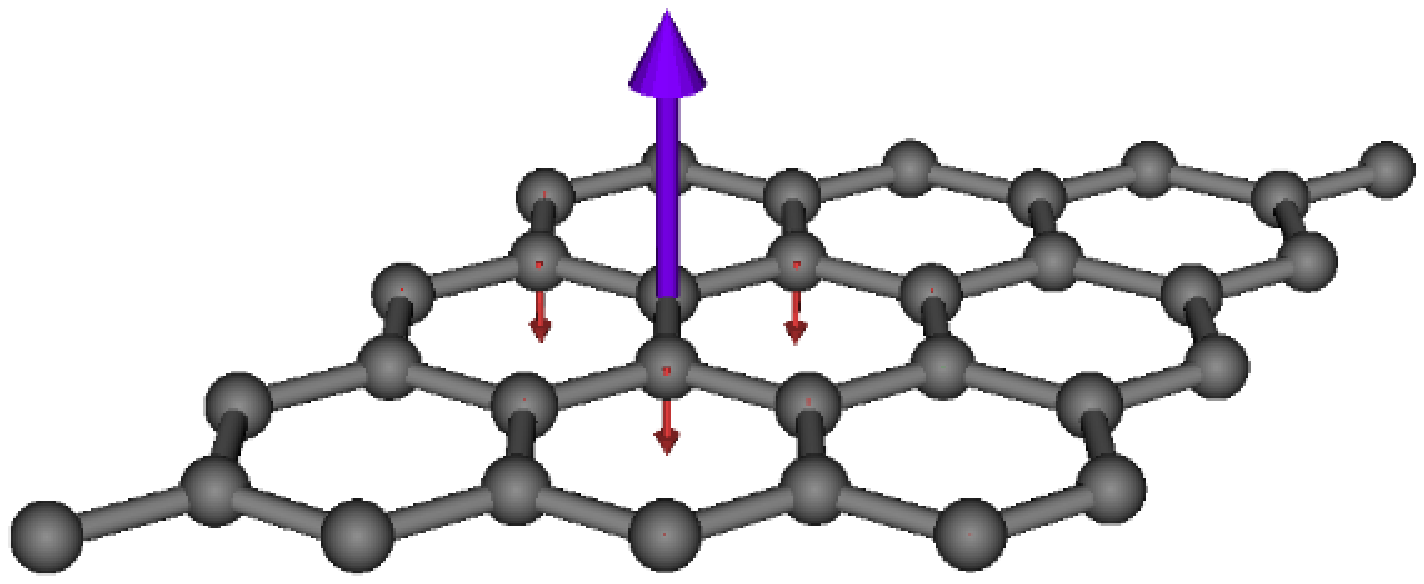}
 \vspace{-20pt}
 \caption{\footnotesize Spin correlation function 
 $\braket{\Psi_{\mathrm{CPS}}|{\bf S}_{i} \cdot {\bf S}_{j}|\Psi_{\mathrm{CPS}}}$ of Graphene in the $p_z$ space with a six-site CPS 
 (with $i$ as the atomic site with maximal spin)\cite{HPV:VisIt}. }
 \label{fig:correlation}
 \vspace{-10pt}
\end{figure}

From the sampled density matrix, we can construct the spin correlation function to analyse the extent to which spin fluctuations 
among the $\pi$/$\pi^*$-bands around the Fermi level affect the magnetic order of the system. The spin correlation functions are constructed 
from two-point functions, rather than from symmetry-breaking in the wavefunction, and show
a rapid decay of anti-ferromagnetic correlations which only substantially affect nearest neighbours (fig.~\ref{fig:correlation}). 

\textit{Conclusions.--} In this work we have presented a novel approach to sample and optimize arbitrary 
non-linear wavefunctions of many-body quantum systems. The optimization is written as an accelerated propagator 
inspired by ideas from developments in deep learning algorithms and the FCIQMC approach. This allows for large 
numbers of parameters to be handled, and systematically improvable Fock-space wavefunctions to be used in both 
lattice and {\it ab initio} systems. 

\begin{acknowledgments}
\textit{Acknowledgments --}
G.H.B. gratefully acknowledges funding from the Royal Society via a University Research Fellowship, as well as the support from the Air Force Office of Scientific
Research via grant number
FA9550-16-1-0256. A.A. has been supported by 
EPSRC, grant number: EP/J003867/1. L.R.S. is supported by an EPSRC studentship. 
The calculations made extensive use of computing facilities of the Rechenzentrum Garching of the Max 
Planck Society with research datasets from this work available at https://doi.org/10.17863/CAM.8161.
\end{acknowledgments}

%

\end{document}